\documentstyle[aps,graphicx]{revtex}
\begin{document}
\title{Isospin and F-spin structure in low-lying levels of  $^{48,50}$Cr isotopes }
\author{ F.H. Al-Khudair$^{1,2}$  and  G.L. Long $^{1,2,3,4}$ }
\address{$^{1}$Department of Physics,  Tsinghua University, Beijing 100084, China\\
$^{2}$Key Laboratory for Quantum Information and Measurements, MOE
100084, China\\
$^{3}$Center of Nuclear theory, Lanzhou Heavy Ion Accelerator
National Laboratory, Lanzhou, 730000, China.\\
$^{4}$Institute of Theoretical Physics, Chinese Academy of
Sciences, Beijing, 100080, China}
\maketitle
\date{{\it Keywords}: IBM-3, Isospin, Mixed Symmetry States,
$^{48,50}$Cr isotopes }

\begin{abstract}
 The low-energy level structure and electromagnetic transitions
of  $^{48,50}$Cr nuclei have been studied by using interacting
boson model with isospin (IBM-3). A sequence of isospin
excitation  bands  with isospin $(T= T_{z}$, $T_{z}+1$ and $
T_{z}+2 ) $ has been assigned, and compared with available data.
According to this study, the $2^{+}_{3} $ and $2^{+}_{2} $ states
are the lowest mixed symmetry states in  $^{48,50}$Cr
respectively. In particular, the present calculations suggest
that a combination of isospin and F-spin excitation can explain
the  structure in these nuclei. The transition probabilities
between the levels are analyzed in terms of isoscalar and
isovector decomposition which reveal the detailed nature of the
energy levels. The results obtained are found in  good agreement
with  recent experimental data.
\end{abstract}

\section{Introduction}

The interacting boson model (IBM) \cite{Ari1,Ari2,Ari3} is
capable of giving  a simple yet realistic description of nuclear
 collective motions.
  In its  original version (IBM-1), only  one kind of boson is
  considered, corresponding to fully proton neutron symmetric
  states. The  neutron-proton  extension of the model ( IBM-2),
   predicts a new class of states \cite{Iac1} having   mixed
   symmetry in the proton and neutron degrees of freedom, and
    it has been successfully observed in  various
   experiments\cite{Ham,Faz,Wie}.

In  lighter nuclei, the valence protons and neutrons fill the same
major shell and isospin should be taken into account. In order to
include the isospin,
 IBM has been extended to the interacting boson model with
isospin (IBM-3)\cite{Ell3}. In IBM-3 three types of  bosons are
included: proton-proton  $(\pi)$, neutron-neutron $(\nu)$ and
proton-neutron $(\delta)$ which forms the isospin $ T=1$
multiplet. The $\nu$, $\delta$ and $\pi$ bosons have the isospin
projection $M_{T}= -1,0,1$ respectively. The wave functions can
be classified by the $U_{c}(3)\supset SU(2)_{T}$ group
\cite{Ell3}, where $SU(2)_{T}$ is the usual isospin group.
Dynamical symmetries of the IBM3 have been studied in
Refs.\cite{Lon1,Kota,Lon2,Lac1,Ginocchio,Kota2}.  The $U_{sd}(6) $
of IBM-1 goes to $ U(18) $ group for IBM-3 as its dynamical
symmetry group. The natural chains below $U(18) $ start with $
U_{sd}(6) \times U_{c}(3) $ , and they must contain $O(3) $ and
$SU_{T}(2)$ as subgroups  because the angular momentum and the
 isospin are good quantum numbers. The chains beginning with
 $ U_{sd}(6) \times U_{c}(3) $  and satisfying the above requirement are the
 following \cite{Lon1}

 \begin{eqnarray}\label{e1}
 U(18)&\supset &(U_{c}(3)\supset SU_{T}(2)) \times (U_{sd}(6) \supset U_{d}(5) \supset
 O_{d}(5) \supset O_{d}(3))\nonumber\\
 U(18)&\supset& (U_{c}(3)\supset SU_{T}(2)) \times (U_{sd}(6) \supset O_{sd}(6) \supset
 O_{d}(5) \supset O_{d}(3))\nonumber\\
 U(18)&\supset& (U_{c}(3)\supset SU_{T}(2)) \times
(U_{sd}(6) \supset SU_{sd}(3) \supset
  O_{d}(3)).
\end{eqnarray}
The subgroups  $ U_{d}(5) $ , $ O_{sd}(6) $ and $ SU_{sd}(3)$
 describe vibrational , $ \gamma $- unstable and rotational
nuclei respectively\cite{Lip}.
 The existence of the isospin excitations in light nuclei
  has received  interest in the last few years\cite{Lan3,Fra,Dur,Fis,Vo2}. Such
   renewed interest was
  sparked by discovery of states with isospin
  $T>T_{z}(T_{z}= \mid Z-N \mid / 2)$. The Cr isotopes have been the
subject of many  theoretical and experimental investigations. On
the experimental side, large $\gamma-$ ray detector arrays
 are now available, and they have increased the detection sensitivity by orders
of magnitude. Meanwhile shell model calculations in full fp shell
can now be performed. Detailed  description of the yrast band
structure in the  full fp space has been reported in $^{48}$Cr
\cite{Cau1} and $^{50}$Cr\cite{Mar1} respectively.
Experimentally, the positive parity yrast band was extended up to
the band termination and turned out to be in  good agreement with
the shell model prediction\cite{Cam1,Lan1}. A detailed analysis
of backbending mechanism of $^{48}$Cr has been done using
projected shell model\cite{Ken1}.
  In $^{50}Cr$ the first backbending at $J^{\pi}=10^{+}$ and second
 backbending at $J^{\pi}=18^{+}$ have been observed experimentally
 \cite{Len1}. The high spin states and electromagnetic
 transitions have been  investigated  and compared with
 the shell model results\cite{Has1}. Recently  microscopic
  three-cluster model  calculations were also performed in an attempt to explain
   the bands structure properties\cite{Des}. However a comprehensive
 analysis of the low-lying levels in these nuclei are still
 missing. It is the  purpose of this paper to study the low-lying
 levels in the IBM-3, in particular to investigate  the isospin and
  F-spin contents of the levels. This  paper is
  divided as follows. In section \ref{s2},
 we briefly discuss the interacting boson model with isospin.
 In section \ref{s3}, we present the results of our calculation
 for the energy levels and compared with available data ,
   and analyze the isospin and F-spin structure of the results. A discussion of  electromagnetic
 transitions follows in section \ref{s4}. Finally, in section \ref{s5} we summarize our results.

  \section{The IBM-3 operators}
\label{s2}

 The most general IBM-3 Hamiltonian can be written as
 \begin{equation}\label{e2}
   H=\epsilon_{s}\hat{n}_{s}+\epsilon_{d}\hat{n}_{d}+ H_{2},
\end{equation}
where
\begin{eqnarray}\label{e3}
H_{2}& =&
\frac{1}{2}\sum_{L_{2}T_{2}}C_{L_{2}T_{2}}((d^{\dagger}d^{\dagger})^{L_{2}T_{2}}.
(\tilde{d}\tilde{d})^{L_{2}T_{2}})
+\frac{1}{2}\sum_{T_{2}}B_{0T_{2}}((s^{\dagger}s^{\dagger})^{0T_{2}}.
(\tilde{s}\tilde{s})^{0T_{2}}) \nonumber\\
&&
+\sum_{T_{2}}A_{2T_{2}}((s^{\dagger}d^{\dagger})^{2T_{2}}.(\tilde{d}\tilde{s})^{2T_{2}})
+\frac{1}{\sqrt2}\sum_{T_{2}}D_{2T_{2}}((s^{\dagger}d^{\dagger})^{2T_{2}}.
(\tilde{d}\tilde{d})^{2T_{2}})\nonumber\\
&&+\frac{1}{2}\sum_{T_{2}}G_{0T_{2}}((s^{\dagger}s^{\dagger})^{0T_{2}}.
(\tilde{d}\tilde{d})^{0T_{2}}),
\end{eqnarray}
and
  \begin{eqnarray}\label{e4}
  (b^{\dagger}_{1}b^{\dagger}_{2})^{L_{2}T_{2}}.
  (\tilde{b_{3}}\tilde{b_{4}})^{L_{2}T_{2}}=(-1)^{(L_{2}+T_{2})}
\sqrt{(2L_{2}+1)(2T_{2}+1}[(b^{\dagger}_{1}b^{\dagger}_{2})^{L_{2}T_{2}}
\times(\tilde{b_{3}}\tilde{b_{4}})^{L_{2}T_{2}}]^{00},
\end{eqnarray}
is the dot product in both angular momentum and isospin. The
tilted quantity is defined as
\begin{equation}\label{e5}
    \tilde{b}_{lm,m_{z}} =(-1)^{(l+m+1+m_{z})}b_{l-m-m_{z}}.
\end{equation}
  The symbols  $T_{2}$ and $ L_{2}$   represent the
 two- boson system isospin and angular momentum.
 The parameters $A$, $B$, $C$, $ D $ and $G$ are related to the
 two-body matrix elements by
 $A_{T_{2}}=\langle sd20\mid H_{2}\mid sd20\rangle$, with $
 T_{2} = 0,1,2 $,  $ B_{T_{2}}=\langle s^{2}0T_{2}\mid H_{2}\mid s^{2}0T_{2}\rangle$,
 $G_{T_{2}}=\langle s^{2}0T_{2}\mid H_{2}\mid d^{2}0T_{2}\rangle$,
 $D_{T_{2}}=\langle sd2T_{2}\mid H_{2}\mid d^{2}2T_{2}\rangle$ and
 $C_{L_{2}T_{2}}=\langle d^{2}L_{2}T_{2}\mid H_{2}\mid d^{2}L_{2}T_{2}\rangle$,
  with $T_{2}= 0,2 $ and $L_{2}= 0,2,4 $ and by
 $C_{L_{2}1}=\langle d^{2}L_{2}1\mid H_{2}\mid d^{2}L_{2}1\rangle$ with $
 L_{2}=1,3$.
  The parameters $A_{1}$, $C_{11}$ and $C_{31} $ are  similar  to
 the Majorana interactions in the IBM-2 which will be referred also as
Majorana interactions. These interactions are important to shift
the states with  mixed symmetry with respect to the total
symmetric ones. Since  only a little experimental information is
known about such states in the nuclei under study, we attempt to
vary   the parameters appearing in these terms to fit the energy
of  available experimental data on $1^{+}$ and other state which
  are sensitive to  these parameters.

The  values of parameters were chosen according to the microscopic
studies in Ref \cite{Eva1}.  The best fit to the whole spectrum
can be found. We have rewritten the Hamiltonian in terms of
linear combination of Casimir operators which is convenient to
analyze the dynamical symmetry nature. The expressions of the
Casimir operators can be found in Ref.\cite{Lon1}. In Casimir
operator form, the Hamiltonians are
\begin{eqnarray}\label{e6}
  H_{48} &=&   \lambda C_{2U_{sd}(6)}+ 1.460 T(T+1)+ 0.030C_{1U_{d}(5)}
   -0.091C_{2SU_{sd}(3)}\nonumber\\
   &&+ 0.025C_{2U_{d}(5)} + 0.173C_{2O_{d}(5)}+0.01C_{O_{d}(3)},\\
  H_{50}& =&  \lambda C_{2U_{sd}(6)}+ 1.322 T(T+1)+ 0.780C_{1U_{d}(5)}
   -0.119C_{2SU_{sd}(3)}\nonumber\\
   &&+ 0.030C_{2U_{d}(5)} + 0.091C_{2O_{d}(5)}+0.002C_{O_{d}(3)},
   \label{e7}
  \end{eqnarray}
for $^{48,50}$Cr  isotopes respectively. The corresponding
parameters in the form of eq.(\ref{e3}) are also given in Table
\ref{t1}.
 The $\lambda $ determines
the position of the mixed symmetry states. In $^{50}$Cr, there is
one $ 1^{+} $ state at 3.629 MeV, and this requires $\lambda
=-0.03$ MeV. We then use the same value for $ ^{48}$Cr  isotope
because no experimental information is available in this isotope.
It can be seen from the Casimir operator form that the
Hamiltonian of $^{48}$Cr is more rotational than that of
$^{50}$Cr because the coefficient of $C_{U_{d(5)}}$ in $^{48}$Cr
is much less than that in $^{50}$Cr.   In $^{50}$Cr  the
coefficient of   $ C_{1U_{d}(5)}$ is very large, and this
indicates that it is more close to the $U(5)$ limit and is in
transition from $U(5)$ to $SU(3)$.

\section{Isospin and F- spin symmetry structure }
\label{s3}

\begin{figure}[htp]
 \begin{center}
\includegraphics[width=6.5in,height=4.5in]{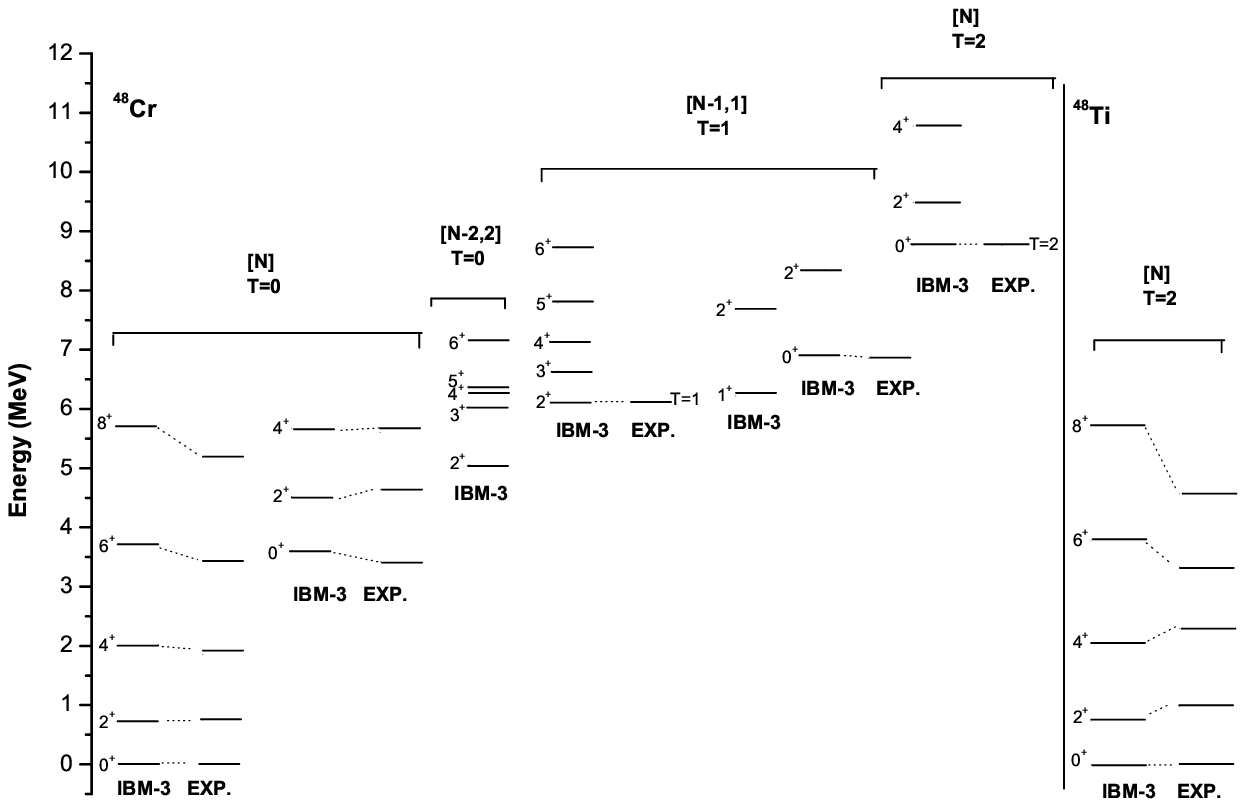}
 \caption{ Comparison between  lowest excitation energy  bands
  $(T=T_{z},T_{z}+1$ and  $T_{z}+2)$ of the IBM-3 calculation
    and experimental excitation energies  of $^{48}Cr$.
  The experimental data are taken from
  Ref.\protect\cite{Ric1}.}\label{f1}
  \end{center}
\end{figure}

\begin{figure}[htp]
\begin{center}
\includegraphics[width=6.5in,height=4.5in]{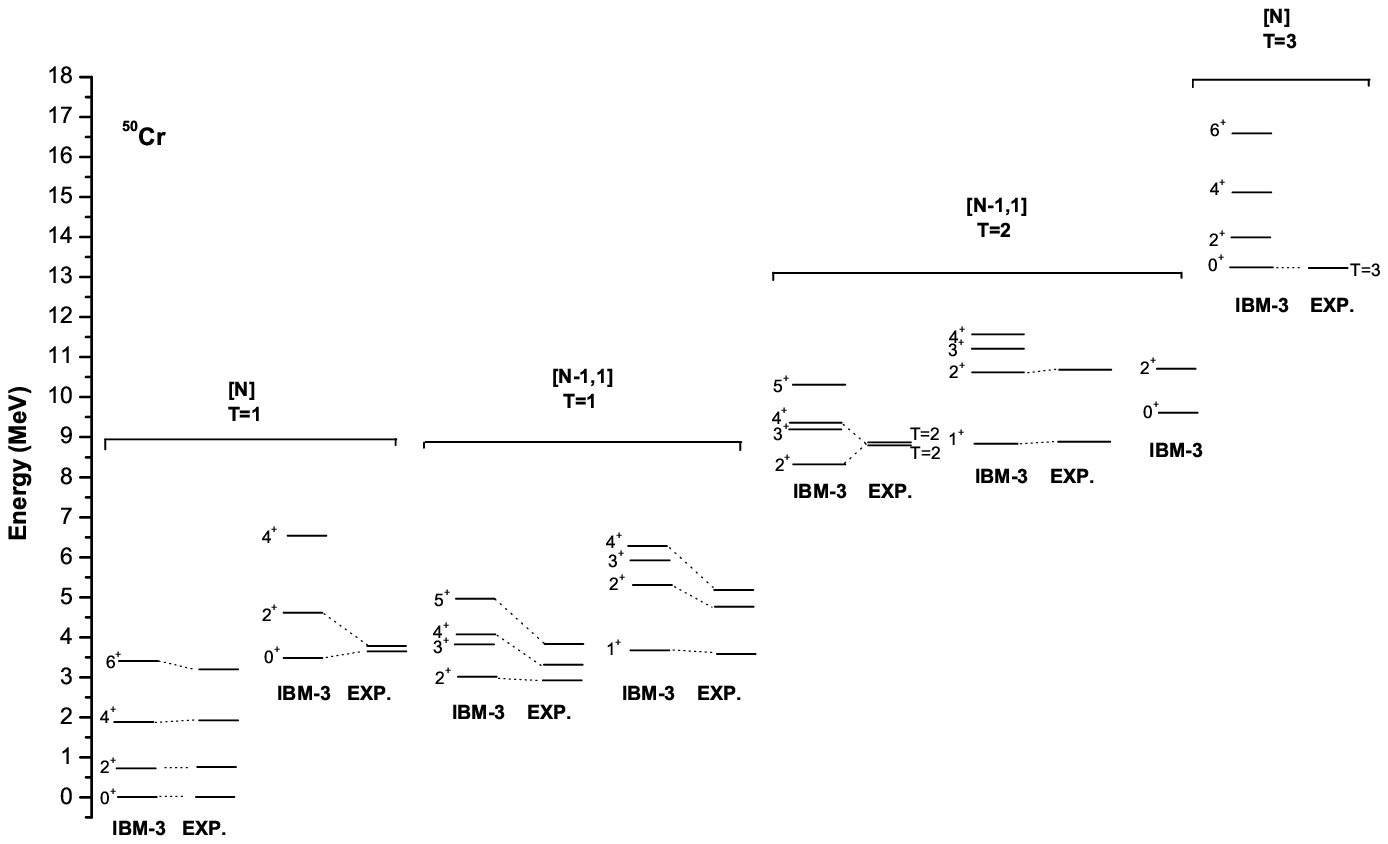}
 \caption{ Comparison between  lowest excitation energy  bands
  $(T=T_{z},T_{z}+1  $ and  $T_{z}+2) $ of the IBM-3 calculation
    and experimental excitation energies  of $^{50}$Cr.
  The experimental data are taken from
  Ref\protect\cite{Ric1}.}\label{f2}
  \end{center}
\end{figure}

 We assume$^{56}$Ni  as the closed core and the bosons in the current study
corresponding to  pairs of hole fermions. The energy levels are
shown in Figs. \ref{f1} and \ref{f2}. Good agreement between the
calculated and observed spectrum is confirmed up to $J\leq 2N$.
The energy levels and wave-functions are obtained using a
computer program written by Van Isacker \cite{Isa2}. It is very
interesting to see that the sequence of bands with $ (T= T_{z}$, $
T_{z}+1$ and $ T_{z}+2 ) $ has been  produced, and is in a good
agreement with available experimental data. The $\beta$- band
with $ T=T_{z} $ is the first and second  excited band in $
^{48}$Cr and $^{50}$Cr respectively. In Fig.\ref{f1} we also draw
the ground state band  of $ ^{48}$Ti where the lowest isospin is
$T=2$ which should be close to the isospin analogue state in $
^{48}$Cr. Here we have assumed that the ground state energy of $
^{48}$ Ti is equal to that of the IBM3 calculated $0^{+}_{T=2}$
state in $^{48}$Cr. This is supported by the following estimate.
We estimate the energy of the isospin analogue state in $^{48}$Cr
by considering the binding energy difference of $^{48}$Cr and
$^{48}$Ti and then subtracting the Coulomb energy difference. This
estimation is crude because Coulomb energy depends on the shape
of the nucleus sensitively. By using the tables in
Ref.\cite{Audi}and the following Coulomb energy formula
\begin{equation}\label{e8}
    E_{Coulmob}= 0.70\frac{Z^{2}}{A^{1/3}}(1-0.76 Z^{-2/3}),
\end{equation}
we obtained the energy of the $T=2$ isospin analogue state in
$^{48}$Cr to be  $ 8.350 $ MeV which is close to the energy of
the $ 0^{+}_{T=2} =8.760$ MeV in our IBM-3 calculation.

 The energy levels in those figures show that there is good
 agreement in the ground state and $\beta $ bands in general.
 The following points need special attention.
 In $^{48}$Cr, the $3^{+}_{1}(T=0) $  state appears at 6.188 MeV in
 our calculation, and it is not yet seen in experiment. In $^{50}$Cr the
 IBM-3 predicts the first $ 3^{+}$ at 3.823 MeV and the second
 $3^{+} $ at 5.921 MeV. These states are both from the $[N-1,1]$ $U(6)$
 irreducible representation. Here we have the $U(6)$ labeling as it is a good quantum number
 approximately. The first and second $J=4^{+}$
  in $^{48,50}$Cr are in good agreement with experimental data,
   and  $4^{+}_{3}$ state in $^{48}$Cr and  $4^{+}_{2}$ state in $^{50}$Cr are mixed symmetry
   states. In both nuclei the
    first $ J=5^{+}$ state is a solely mixed symmetry state.

\begin{figure}[htp]
\begin{center}
\includegraphics[width=5in,height=4in]{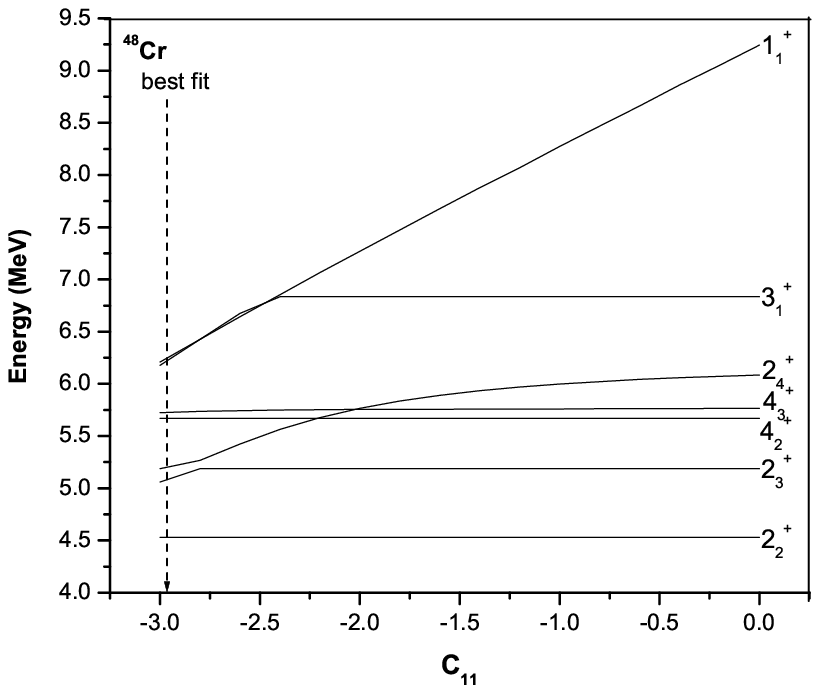}
\caption { The variation in level  energy of $^{48}Cr $ as a
function of $C_{11}$; all the other parameters were kept at their
best-fit values.}\label{f3}
\end{center}
\end{figure}

\begin{figure}[htp]
\begin{center}
\includegraphics[width=5in,height=4in]{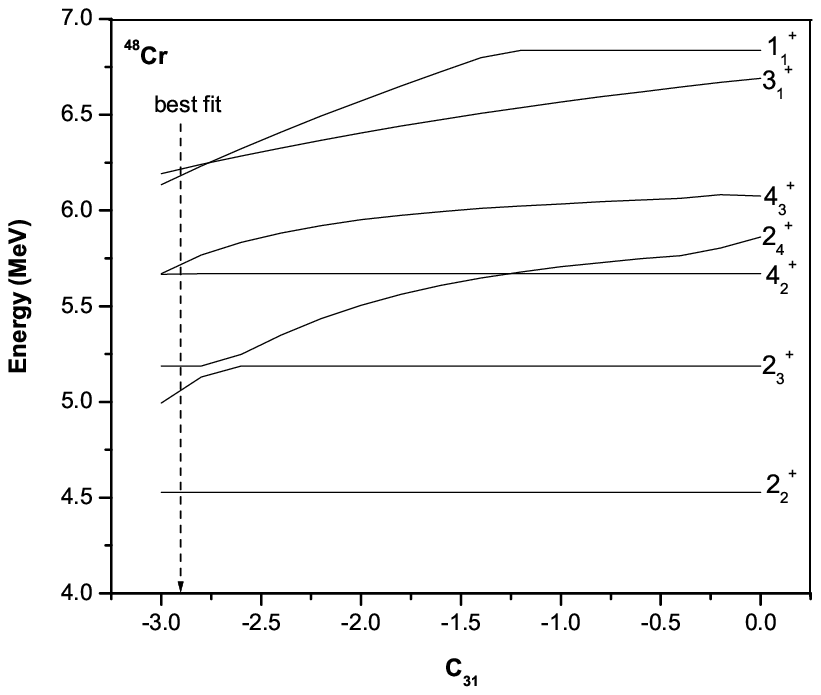}
\caption { The variation in level  energy of $^{48}Cr $ as a
function of $C_{31}$; all the other parameters were kept at their
best-fit values.}\label{f4}
\end{center}
\end{figure}

\begin{figure}[htp]
\begin{center}
\includegraphics[width=5in,height=4in]{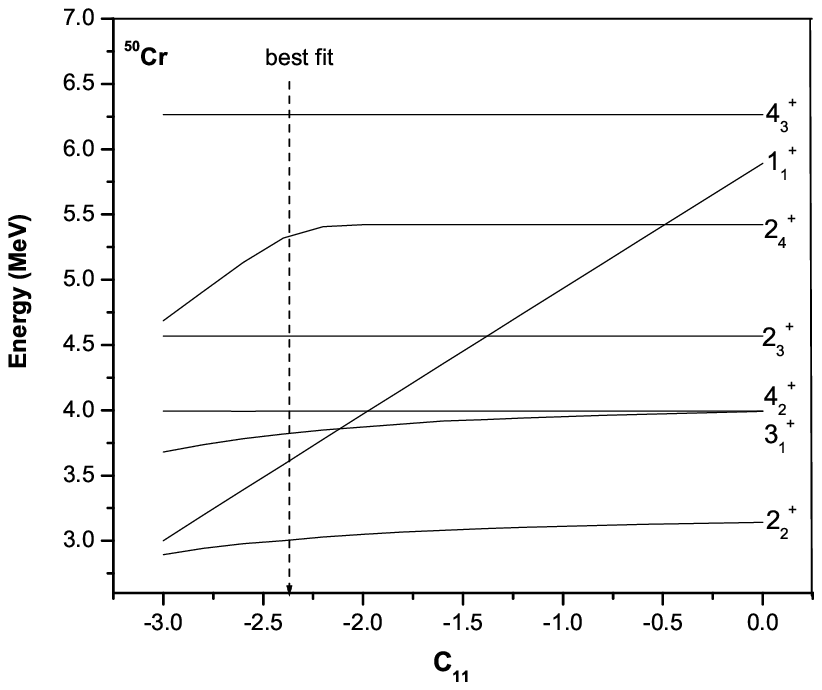}
\vskip -0.2cm \caption { The variation in level  energy of
$^{50}Cr $ as a function of $C_{11}$; all the other parameters
were kept at their best-fit values.}\label{f5}
\end{center}
\end{figure}

\begin{figure}[htp]
\begin{center}
\includegraphics[width=5in,height=4in]{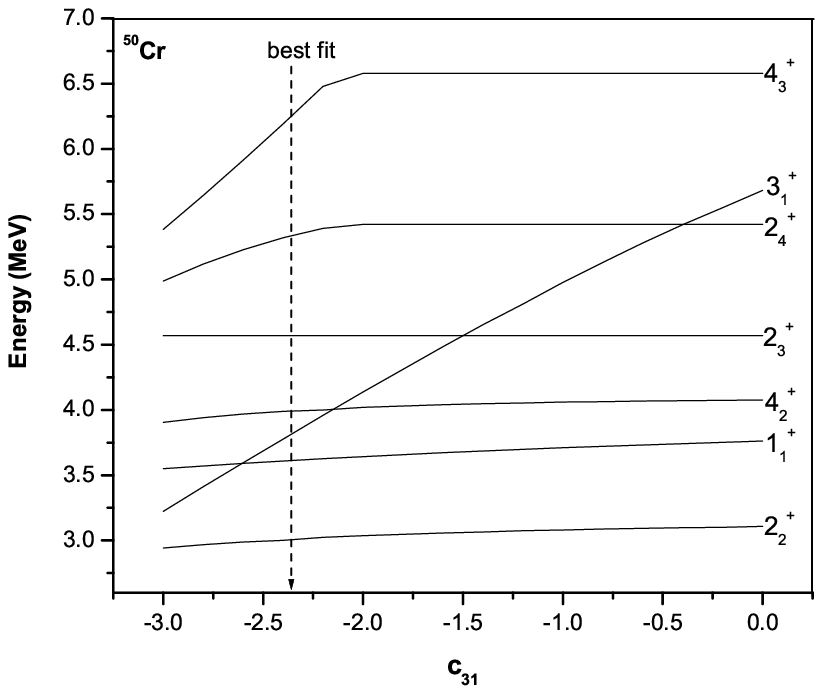}
\caption { The variation in level  energy of $^{50}Cr $ as a
function of $C_{31}$; all the other parameters were kept at their
best-fit values.}\label{f6}
\end{center}
\end{figure}

To identify the mixed symmetry states, one can make use of their
electromagnetic transition properties: weakly E2 and strong M1
decay to ground state and first $2^{+} $ state respectively
\cite{Isa1}. Mixed symmetry $J=2^{+}$ state in light nuclei have
been identified in $^{54,56}$Cr, $^{56,58}$Fe\cite{Hit1},
$^{56}$Fe \cite{Eid,Hit2} and $^{64-68}Ge,^{60-66}Zn$ \cite{Ell2}.
The mixed symmetry structure of wave functions can be seen by
calculating the  $\langle J \mid C_{2U(6)}\mid J\rangle$ value.
The calculated $2^{+}_{3}$ and $2^{+}_{2}$ in $^{48}$Cr and
$^{50}$Cr exhibit $[N-1,1]$ $U(6)$ partition, indicating
 that those states are the lowest mixed symmetry states
  in $^{48,50}$Cr respectively.  For $^{48}$Cr($ T_{z}=0 $),
  the lowest mixed symmetry state come from $ [N-2,2] $ because
  $[N-1,1]$ does not contain $T=0$,
 while for $^{50}$Cr  it comes from $ [N-1,1] $.
 From Figs. \ref{f1} and \ref{f2} one can see this fact and  therefore the
 lowest mixed symmetry state in $^{48}$Cr has high energy.
  The IBM-3 analysis gives a $ 1^{+}_{1}$ level at  $6.218$ MeV
with partition $[N-1,1]$, and it is higher than the lowest mixed
symmetry $2^{+}$ state. No experimental evidence for this level
is available. The energy of the first $1^{+}$ in $^{50}$Cr equals
to $3.613$ MeV, and it is quite close to the experimental one. A
possible candidate for $1^{+}_{2}$ mixed symmetry has also been
identified at 7.876 MeV, it is close to the observed state  $ J=(
1,2) $ at $7.646 MeV$. The energy of these observed states  are
well reproduced by the calculation.  We have varied each of
Majorana parameters around the best-value and keeping all other
parameters at their best-fit values, the variations of the energy
of these states with the parameters are shown in  Figs
\ref{f3}-\ref{f6}. In $^{48}$Cr  the mixed symmetry component of
the lowest mixed symmetry band belongs predominantly to the
partition $[N-2,2]$ with $ T= T_{z} $. The second mixed symmetry
state $2^{+}$ has the partition $[N-1,1]$ with $T=T_{z}+1=1$ with
energy 6.110 MeV, it is very close to  experimental level at
6.100 MeV $(T=1)$ \cite{Ric1}. Starting with this mixed state, a
whole band of mixed states is predicted by this IBM3 calculation.
In the same time the lowest mixed symmetry bands in $^{50}$Cr
have the partition $[N-1,1]$ with $ T= T_{z},T_{z}+1  $ as shown
in Fig. \ref{f2}. Because IBM-3 has three charge states, it is
possible to have $U(6)$ partitions into three rows, namely the $
[N_{1},N_{2},N_{3}]$ states which are characteristic of IBM-3. We
found that such state come high in energy, upwards at about 8.5
MeV, and the lowest example being a scissor  mode state  in
$^{48}$Cr at 8.599 MeV which is predominantly the  $ [2,1,1]$
partition with $T=1$. The present calculation predicts that it
 decays to $ 2^{+}_{3}$ by strongly M1 transition with $B(M1)=0.24
\mu_{N}^{2}$ and weakly E2 transition with $B(E2) = 0.0013
e^{2}b^{2} $. It is very significant if these properties are
observed in experiment.

\section {electromagnetic transition}
\label{s4}

 The E2 transition is described by the following isoscalar and isovector one
boson E2 operators $ Q =Q^{0} + Q^{1} $ \cite{Ell1}
\begin{eqnarray}\label{e9}
Q^{0}&=&\alpha_{0}\sqrt{3}[(s^{+}\hat{d})^{20}+(d^{+}\hat{s})^{20}]+
\beta_{0}\sqrt{3}[(d^{+}\hat{d})^{20}\\
\label{e10}
Q^{1}&=&\alpha_{1}\sqrt{2}[(s^{+}\hat{d})^{21}+(d^{+}\hat{s})^{21}]+
\beta_{1}\sqrt{2}[(d^{+}\hat{d})]^{21}.
\end{eqnarray}

 The  M1 transition is also a one boson operator with
  an isoscalar part and an isovector  part $ M= M^{0} +M^{1}
 $\cite{Ell1},
\begin{eqnarray}
M^{0}&=& g_{0} \sqrt{3}(d^{+}\hat{d})^{10}=g_{0}
L/\sqrt{10}\label{e11}\\
M^{1}&=& g_{1} \sqrt{2}(d^{+}\hat{d})^{11}\label{e12},
\end{eqnarray}
where $ g_{1} $ and $g_{0} $ are the isovector and isoscalar
$g$-factor respectively and $ L $ is angular momentum operator.

Having obtained the wave function for states , we can calculate
the electromagnetic transition rate between states using the
subroutine of Lac \cite{Lac2}. In order to link it with energy
code, we have made some modification. The parameters in the E2
operator are adjusted to fit the experimental data of
$B(E2;2^{+}_{1}\rightarrow 0^{+}_{1})$, where
$\alpha_{0}=\beta_{0}= 0.1,0.13 $ for$^{48}$Cr and $^{50}$Cr
respectively and $\alpha_{1}=\beta_{1}= 0.1 $  for both isotopes.
The results are summarized in tables \ref{t2} and \ref{t3}
respectively.

The $B(E2)$ values between the yrast states are well reproduced.
The calculated $B(E2;4^{+}_{1}\rightarrow 2^{+}_{1})$ and
$B(E2;6^{+}_{1}\rightarrow 4^{+}_{1})$ are quite close to the
experimental values, in the same time  the calculated
$B(E2;2^{+}_{3}\rightarrow 0^{+}_{1})$ in $ ^{50}$Cr is larger
than the experimental one by about  $ 0.004 e^{2}b^{2}$. From
these tables one can see that the transitions between states with
the  same partition are larger than transition between states
with different $U(6)$ labels.

 The isovector part $M1$ of the  transition from $ 1^{+}  $ and higher energy $2^{+} $
 mixed symmetry states to ground state and first $2^{+}$ state
 respectively is analyzed. The result is consistent with the
identification of the $1^{+}$ and the lowest $2^{+}_{ms}$ in the
these two nuclei . From the  M1 operator,  the isoscalar part is
simply the angular momentum, M1 transition are determined by
$g_{1} $ part only. For a good reproduction of calculated M1
transition we used the $g_{1} = 2.8$  for the $^{48}$Cr and
$^{50}$Cr . They lead to pure isovecter M1 transitions. In
$^{50}$Cr isotope the calculated $B(M1;1^{+}_{1}\rightarrow
0^{+}_{1}) $ and $B(M1;4^{+}_{2}\rightarrow 4^{+}_{1}) $ values
are in good agreement with the experimental data. We obtain
$B(M1;2^{+}_{2}\rightarrow 2^{+}_{1}) $ =0.6319 $\mu_{N}^{2}$
which compares well with the 0.3938 $\mu_{N}^{2}$ value observed
experimentally. The $ 2^{+}_{2}\rightarrow
 2^{+}_{1} $ decay with its $ E2/M1$ mixing ratio of $\delta $ =
-0.03(6) experimentally  is nearly pure M1 transition.

\begin{figure}[htp]
\begin{center}
\includegraphics[width=6in,height=4in]{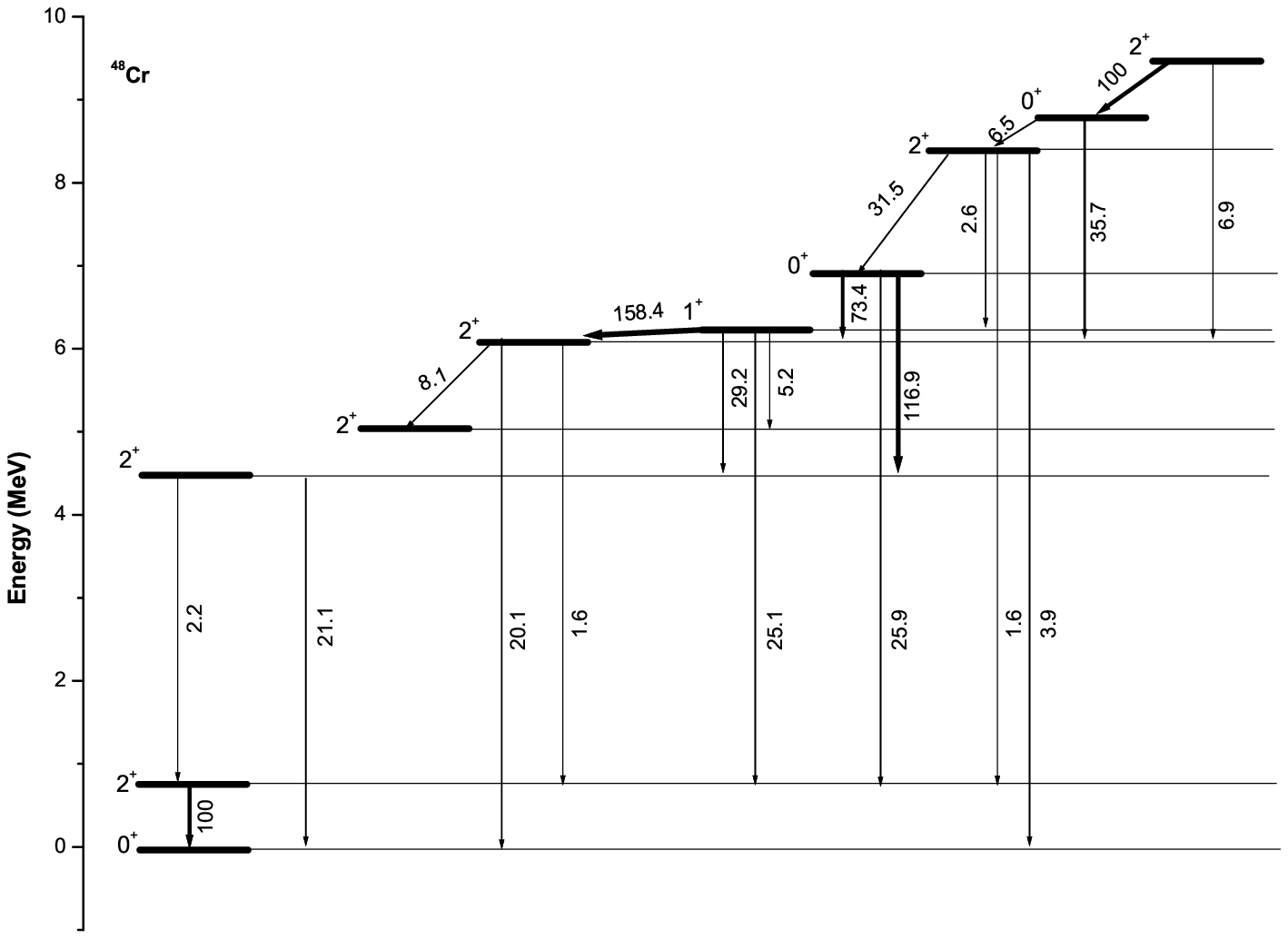}
\caption { Energy levels  and  relative B(E2)  for $^{48}$Cr
isotope.}\label{f7}
\end{center}
\end{figure}

\begin{figure}[htp]
\begin{center}
\includegraphics[width=6in,height=4in]{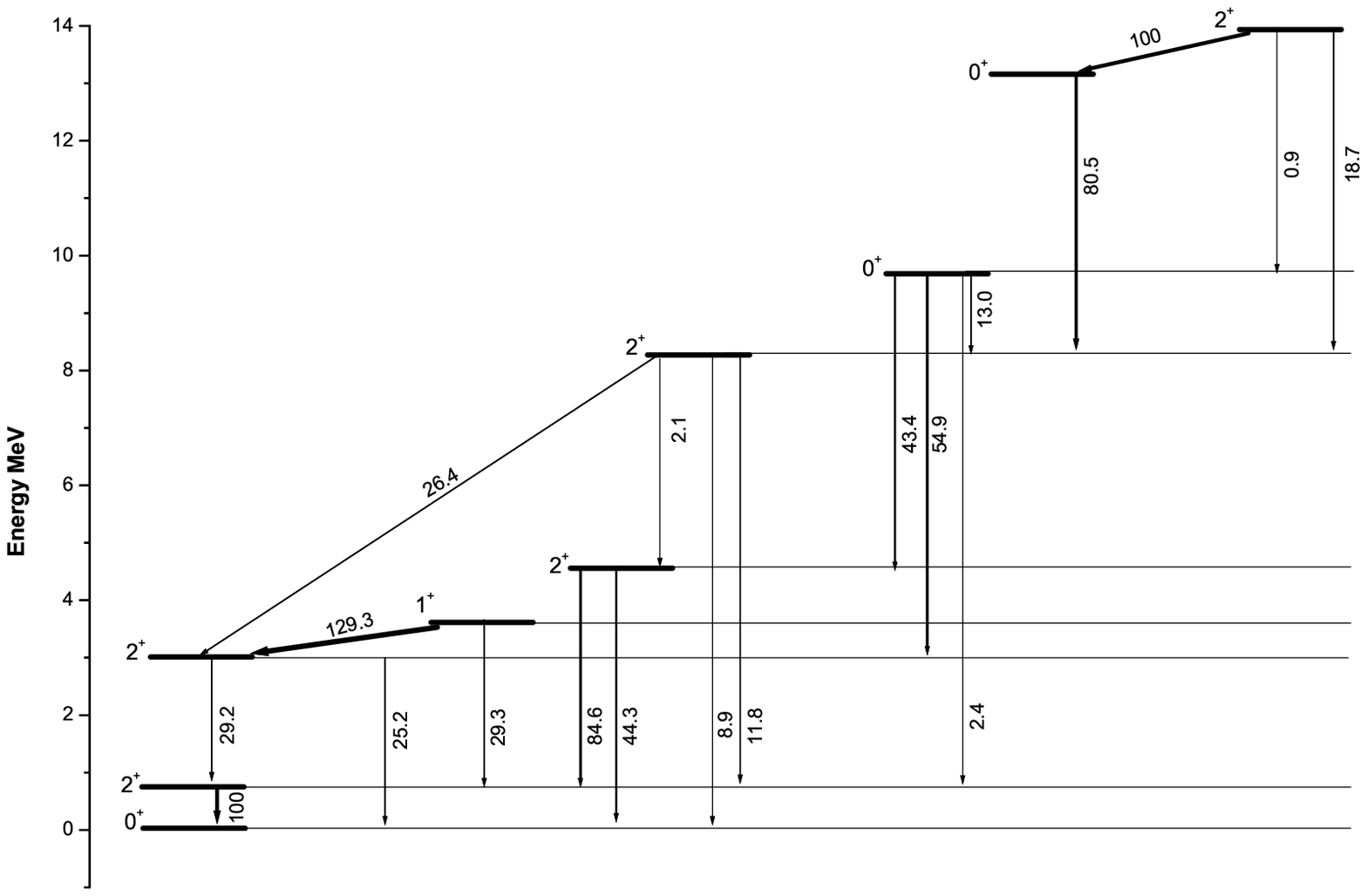}
\caption { Energy levels  and  relative B(E2)  for $^{50}$Cr
isotope. }\label{f8}
\end{center}
\end{figure}

  In $^{50}Cr $ the calculated  $ 1^{+}$ state
  at 8.901 MeV decays to ground state
  with reduced $M1$  transition   0.100 $\mu_{N}^{2}$ and is in
  good agreement  with experimental data of
  0.0823(89)$\mu_{N}^{2}$ reported in
 \cite{Ric1}.
  The relative B(E2)  transitions between  the
states  $  J=0^{+}, 1^{+} $ and  $ 2^{+} $ have  different isospin
values  are calculated and summarized in Figs. \ref{f7} and
\ref{f8} respectively.

\section {Conclusion}
\label{s5}

 Summarizing our results we may conclude that the IBM-3
 description of the low-lying levels in the
$^{48,50}$Cr  nuclei is satisfactory.  We have shown, on the
basis of  energy levels and electromagnetic properties, the
presence of mixed symmetry states near   5  and
  3 MeV   in $^{48,50 }$Cr respectively. In $^{48}$Cr  the
$2^{+}_{3}$ state at 5.069 MeV decays predominantly  to the
$2^{+}_{1} $ state via a pure M1 transition, while in $^{50}$Cr
the $2^{+}_{2}$ state decays through  strongly M1 and weakly
though E2 transition decay to $2^{+}_{1} $. Theses states are the
candidate lowest mixed symmetry states in $^{48,50}$Cr
respectively. In our calculation the scissor mode $1^{+}$ states
in these nuclei lie higher than the lowest mixed symmetry $2^+$
states , which were observed at 3.629 MeV in $^{50}$Cr. No mixed
symmetry state labeled with $[N-1,1]$ and $T=0$ exists in
$^{48}$Cr, and this is a natural result of group reduction. As a
result, the scissor state in $^{48}$Cr nucleus is higher than
that in $^{50}$Cr.

The calculated results are in good agreement with  available
experimental data, but more experiment data for those nuclei are
needed  to validate this nuclear structure predicted. In
particular, the present calculations suggest a combined isospin
and mixed symmetry excitation in the low-lying levels of these
nuclei. The present calculation also predicts the existence of
row-rowed partition state at about 9 Mev, and this type of states
is typical of IBM3, not present in IBM1 and IBM2. It will be
highly desirable to substantiate these predictions in future
experiment.

 Work is supported in part by National Natural Science  Foundation of China under
Grant No. 10047001, Major State Basic Research Development Program
under Contract No. G200077400

\begin{table}
\begin{center}
\caption{ The parameters of the IBM-3 Hamiltonian used for the
description of the Cr-isotopes.}
\label{t1}
\begin{tabular}{lll}\hline
   Nucleus & $^{48}$Cr & $^{50}$Cr \\  \hline
  $\epsilon_{s\nu} = \epsilon_{s\pi}$& 1.830 & 1.274 \\
  $\epsilon_{d\nu} = \epsilon_{d\pi}$ & 2.737 & 2.58 \\
  $A_{i}(i=0,1,2)$& $-$6.264,$-$2.496,2.496 & $-$5.824,$-$2.108,2.108  \\
  $C_{i0}(i=0,2,4)$ & $-$7.172,$-$5.018,$-$6.152 & $-$5.802,$-$4.404,$-$6.042  \\
  $C_{i2}(i=0,2,4)$ & 1.588,3.742,2.608 &2.130,3.528,1.890    \\
  $C_{i1}(i=1,3)$ & $-$2.993,$-$2.892 & $-$2.370,$-$2.350  \\
  $B_{i}(i=0,2)$ & $-$5.900,2.860 &  $-$5.348,2.584 \\
  $D_{i}(i=0,2)$& 0.681,$-$0.814 & 0.890,$-$1.064 \\
  $ G_{i}(i=0,2)$& $-$0.814,2.496& $-$1.064,0.890\\ \hline
\end{tabular}
\end{center}
\end{table}

\begin{table}
\begin{center}
\caption{Experimental\protect\cite{Has1,Ric1} and calculated B(E2)
($e^{2}B^{2}$), calculated  B(M1) ($\mu_{N}^{2}$) and mixing ratio
for $^{48}$Cr isotope.}\label{t2}
\begin{tabular}{lllllll}\hline
    &   $B(E2)$ & & &  $B(M1)$&  & $\delta$  \\
  $J_{i}^{+}\rightarrow  J_{f}^{+}$& Exp. & Calc. &  &Calc. & &Calc.\\  \hline
$2^{+}_{1}\rightarrow0^{+}_{1}$ & 0.0320(41)& 0.0309 &  &  \\
$2^{+}_{2}\rightarrow0^{+}_{1}$ &  & 0.0067&  &  &  & \\
$2^{+}_{2}\rightarrow2^{+}_{1}$ &  & 0.0007&  &0.0000  &  & \\
$2^{+}_{2}\rightarrow0^{+}_{2}$ &  & 0.0101&  &  &  & \\
$1^{+}_{1}\rightarrow0^{+}_{1}$ &   &   &  & 0.6437 &  & \\
$1^{+}_{1}\rightarrow0^{+}_{2}$ &   &   &  & 0.2362 &  & \\
$1^{+}_{1}\rightarrow2^{+}_{1}$ &   & 0.0077 &  & 0.3923 &  & -0.3141\\
$1^{+}_{1}\rightarrow2^{+}_{2}$ &   &0.0091   &  & 0.2103&  &0.1434\\
$1^{+}_{1}\rightarrow2^{+}_{3}$ &   &0.0016   &  & 0.1365&  &0.0517\\
$2^{+}_{2}\rightarrow4^{+}_{1}$ &  & 0.0166&  &  &  & \\
$4^{+}_{1}\rightarrow2^{+}_{1}$ & 0.0329(110) & 0.0383&  &  &  & \\
$4^{+}_{2}\rightarrow2^{+}_{1}$ &   & 0.0012&  &  &  & \\
$6_{1}^{+}\rightarrow4^{+}_{1}$ &0.0301(78) &0.0345& &  &  &
\end{tabular}
\end{center}
\end{table}

\begin{table}
\begin{center}
\caption{Experimental\protect\cite{Has1,Lang,Ric1} and calculated
B(E2) (in unit $e^{2}b^{2}$) , B(M1) (in unit $\mu_{N}^{2}) $ and
mixing ratio for $^{50}$Cr isotope.}\label{t3}
\begin{tabular}{lllllll}\hline
    &  $ B(E2)$ &   &    $B(M1)$&  &   $\delta$ &   \\
  $J_{i}^{+}\rightarrow  J_{f}^{+}$& Exp. & Calc. & Exp. & Calc.& Exp. & Calc.\\  \hline
  $2^{+}_{1}\rightarrow0^{+}_{1}$ & 0.0217(25) & 0.0247 &  &  &  & \\
  $2^{+}_{2}\rightarrow2^{+}_{1}$ &  & 0.0072 &0.3938(716) &0.6319& $ 0.03^{+0.06}_{-0.04}$ &0.1383 \\
  $2^{+}_{2}\rightarrow0^{+}_{1}$ & 0.0023(8)   & 0.0062 &  &  &  & \\
  $0^{+}_{2}\rightarrow2^{+}_{2}$&    & 0.0385 &  &  &  & \\
  $2^{+}_{2}\rightarrow2^{+}_{2}$ &    & 0.0013 &  & 0.0238 &  & $-$0.1291\\
  $2^{+}_{2}\rightarrow0^{+}_{1}$ &    & 0.0109&  &  &  & \\
  $2^{+}_{3}\rightarrow0^{+}_{1}$ &    & 0.0109&  &  &  & \\
  $2^{+}_{3}\rightarrow2^{+}_{1}$ &    & 0.0209 &  & 0.0000 &  & \\
  $1^{+}_{1}\rightarrow0^{+}_{1}$ &    &   & 0.3300(400) & 0.2778 &  & \\
  $1^{+}_{1}\rightarrow0^{+}_{2}$ &    &   &  & 0.0111 &  & \\
  $1^{+}_{1}\rightarrow2^{+}_{1}$ &    & 0.0072 &  & 0.2225 &  & $-$0.2113\\
  $1^{+}_{1}\rightarrow2^{+}_{2}$&    & 0.0318 &  & 0.0013 &  &1.2259 \\
  $1^{+}_{1}\rightarrow2^{+}_{2}$&    & 0.0086 &  & 0.6559 &  & 0.0447\\
  $1^{+}_{2}\rightarrow2^{+}_{1}$&    & 0.0002 &  & 0.0026 &  & 0.2118\\
  $1^{+}_{2}\rightarrow2^{+}_{2}$&    & 0.0048 &  & 0.0012 &  &4.0168 \\
  $3^{+}_{1}\rightarrow2^{+}_{1}$&    & 0.1421 &  & 0.0221 &  & 0.4452\\
  $3^{+}_{1}\rightarrow4^{+}_{1}$&    & 0.0031 &  & 0.4741&  &0.0636 \\
  $4^{+}_{1}\rightarrow2^{+}_{1}$& 0.0204(25)   & 0.0256 &  &  &  & \\
  $4^{+}_{2}\rightarrow4^{+}_{1}$&    & 0.0110 & 0.1325(375) &0.4885 & $-0.02^{+0.16}_{-0.52}$ & 0.1281\\
  $5^{+}_{1}\rightarrow3^{+}_{1}$&    & 0.0031 &  & &  & \\
  $5^{+}_{1}\rightarrow4^{+}_{1}$& 0.0001   & 0.0171 &  &0.0033 &$-$0.47(16)  &0.5005 \\
  $6^{+}_{1}\rightarrow4^{+}_{1}$& 0.0235(47)   & 0.0165 &  &  &  & \\
\end{tabular}
\end{center}
\end{table}




\begin{thebibliography}{9}
\bibitem {Ari1} A. Arima and F. Iachello, Ann. Phys. (N.Y) {\bf
99}(1976)253.
\bibitem{Ari2} A. Arima and F. Iachello, Ann. Phys. (N.Y) {\bf
111}(1978)201.
\bibitem{Ari3} A. Arima and F. Iachello, Ann. Phys. (N.Y) {\bf
123}(1979)468.
\bibitem{Iac1}F. Iachello and A. Arima, {\it The Interacting
 Boson Model}, Cambrideg University Press, Cambrideg, England
 (1987).
  \bibitem{Ham} W.D. Hamilton, A. Irb\"{a}ck and J.P Elliott, Phys Rev. Lett. {\bf 53} 2469(1984).
\bibitem{Faz}B. Fazekas, T. Belgya, G. Molnar, A. Veres,R.A. Gatenby, S.W. Yates
 and T. Otsuka , Nucl. Phys. {\bf A 548} (1992)249.

\bibitem{Wie}I. Wiedenhover, A. Gelberg, T. Otsuka, N. Pietralla, J. Gableske, A. Dewald
 and  P.von Brentano. Phys. Rev. {\bf C56}( 1997)R2354.

\bibitem{Ell3}J.P.  Elliott and A.P White, Phys. Lett {\bf B 97}(1980)169.
\bibitem{Lon1} G.L. Long Chinese Journal of Nuclear Physics, {\bf
16}(1994)331.
\bibitem{Ginocchio} J. N. Ginocchio, Phys. Rev. Lett. 77 (1996)
28.
\bibitem{Kota}V.K.B. Kota, Phys. Lett {\bf B 399}(1997)185.
\bibitem{Lac1}V.S. Lac, J.P. Elliott  and J.A. Evans, Phys. Lett. {\bf B 394}
 (1997)231.
 \bibitem{Lon2} G. L. Long, C. Y. Gan and T. Y. Shen, Commun.
 Theor. Phys. {\bf 27} (1997) 317.
 \bibitem{Kota2} V.K.B. Kota, Ann. Phys. {\bf 265} (1998) 101.
\bibitem{Lip}P.O. Lipas, International Review of Nucl. Phys.
 Vol.{ \bf 2 }1984, Edited by T. England, J. Rekestad and J-S Vadgen.

 \bibitem{Lan3}K. Langanke, P. Vogel, Dao-Chen Zhen, Nucl. Phys.
{\bf A 626} (1997)735.
\bibitem{Fra} S. Frauendorf and J. Sheikh, Nucl. Phys. {\bf A 645} (1999)509.
 \bibitem {Dur} Y. Durga Devi, Shadow Robmson, and Larry Zamick,Phys. Rev. {\bf C 61}
 (2000)037305.
 \bibitem {Fis} S.M. Fischer $\it{et al.}$, Phys. Rev.
 Lett.{\bf 87}(2002)132501.
 \bibitem{Vo2}P. Von Brentano, C. Frie$\beta$ner, R.N. Jolos,
  A.F.Lisetskiy, A.Schmidt, I.Schneider, N. Pietralla, T. Sebe and T. otsuka
   Nucl. Phys.{\bf A 704} (2002)115c.
\bibitem{Cau1}E. Caurier ,J.L. Edido, G-Martinez-Pinedo, A.poves,J. Retamosa,
L.M. Robledo and A.P. Zuker, Phys. Rev.Lett. {\bf 75}(1995)2466.
\bibitem{Mar1}G. Martinez-Pinedo A.Poves, L.M. Robledo, E.Caurier, F.Nowacki,
 J.Retamosa and A. Zuker, Phys. Rev.{\bf C 54}(1996)R2150.
\bibitem{Cam1} J.A. Cameron G. ,J.L.Rodriguez,J. jonkman, G Hackman,
 S.M. Mullins, C.E. Svensson, J. C. Waddington, Lihong Yao, T.E. Drake, M.Cromaz,
  J. H. DeGraaf and G. Zwartz, H. R Andrews, G.ball, A. Galindo-Uribarri
  , V.P. Janzen, D. C. Radford and D. Ward,  Phys. Rev.{\bf C 58}(1998)808.
\bibitem{Lan1}K. Langanke, Phys. Letter. {\bf B 438}(1998)235.
\bibitem{Ken1}Kenji Hara, Yang Sun and Takahiro Mizusaki, Phys.
Rev. Lett. {\bf 83}(1999)1922.
\bibitem{Len1}S.M. Lenzi, D.R. Nagarajan, D. Bazzacco, D.M.Brink, M.A. Cardona,
, G. de Angelis, M. de Poli, A.Gadea, D.Hojman, S. Lunardi,
N.H.Medina, and C. Rossi Alvarez, Phys. Rev. {\bf C 56}(1997)1313.
\bibitem{Has1} M. Hasegawa , K. Kaneko and S.Tazaki, Nucl. Phys.  {\bf A 674}(2000)411.
\bibitem{Des}P.Descouvemont, Nucl. Phys. {\bf A 709} (2002)275.
\bibitem{Eva2}J.A. Evans, G.L. Long, J.P. Elliott,  Nucl. Phys.
  {\bf A 561}(1993)201.
\bibitem{Eva1}J.A. Evans, J.P. Elliott,  V.S. Lac and G.L. Long, Nucl. Phys.
  {\bf A 593}(1995)85.
\bibitem{Isa2}P.Van Isacker, unpublished.
\bibitem{Audi}G. Audi and A. H. Wapstra, Nucl. Phys. {\bf A 565} (1993)1.
\bibitem{Isa1}P.Van Isacker, K. Heyde, J. Jolie and A. Sevrin
 Ann.Phys.(N.Y) {\bf 171}( 1986)253.
 \bibitem{Hit1} Hitoshi Nakada and Takaharu Otsuka, Phys. Rev. {\bf C 55}
 (1997)2418.
\bibitem{Eid}S.A.A. Eid, W.D. Hamilton and J.P. Elliott, Phys. Lett.
 {\bf B 166}( 1986)267.
\bibitem {Hit2} Hitoshi Nakada, Takaharu Otsuka and Takashi Sebe, Phys. Rev.
 Lett.{\bf 67}(1991)1086.
\bibitem{Ell2}J.P.  Elliott, J.A. Evans, V.S. Lac and G.L. Long, Nucl. Phys.
  {\bf A 609}(1996)1.
 \bibitem{Ell1}J.P. Elliott, J.A. Evans and V.S. Lac, Nucl. Phys. {\bf A 597}
 (1996)341.
  \bibitem{Lac2}V.S. Lac, unpublished (1996).
  \bibitem{Lang}J. Lang, Krishna Kumar and J. H. Hamilton, Rev. of Mod. Phys.,
   Vol.{\bf 54}, No. 1(1982) 119.
\bibitem{Ric1}Richard B. Firestone, {\it
Table of Isotopes}, Eighth Edition, edited by  Virginia S. shirley
1998.
 \end{thebibliography}
\end{document}